\documentclass{iau}
\usepackage{natbib,graphicx}
\def\msun{{\rm\,M_\odot}}

\newcommand{\aap}{\textit{A\&A}}
\newcommand{\aaps}{\textit{A\&AS}}

\newcommand{\aj}{\textit{AJ}}

\newcommand{\apj}{\textit{ApJ}}
\newcommand{\apjl}{\textit{ApJ}}

\newcommand{\araa}{\textit{ARA\&A}}

\newcommand{\mnras}{\textit{MNRAS}}
\newcommand{\nat}{\textit{Nat}}

\title[Dark matter in massive galaxies] 
{Dark matter in massive galaxies}

\author[Ortwin Gerhard]   
{Ortwin Gerhard$^1$}

\affiliation{$^1$Max-Planck-Institut f\"ur extraterrestrische Physik, \\ 
Postfach 1312, Giessenbachstr., 85741 Garching, Germany \\
email: {\tt gerhard@mpe.mpg.de}}

\pubyear{2013}
\volume{295}  
\pagerange{xx--xx}
\jname{The intriguing life of massive galaxies}
\editors{D. Thomas, A. Pasquali \& I. Ferreras, eds.}
\begin{document}

\maketitle

\begin{abstract}
  The spatial distributions of luminous and dark matter in massive
  early-type galaxies reflect the formation processes which
  shaped these systems. This article reviews the predictions of
  cosmological simulations for the dark and baryonic components of
  ETGs, and the observational constraints from lensing, hydrostatic
  X-ray gas athmospheres, and outer halo stellar dynamics.
\end{abstract}

\firstsection 
\section{Introduction}

A galaxy's dark matter mass is arguably the most important parameter
characterizing its evolutionary state. Massive early-type galaxies (ETGs)
with virial masses of up to $\sim 10^{13}\msun$ are those that
have evolved furthest in the galaxy building process, with little gas
left to make new stars.  In the centers of ETGs the luminous matter
dominates, but already at a few half-light radii ($R_e$) of the
stellar distribution, dark matter (DM) takes over. The spatial
distributions of luminous and dark matter in the inner regions of ETGs
are the outcome of the formation processes by which these systems were
shaped, and are thus one of the observational constraints we have on
these processes, next to scaling relations, stellar population
properties, and orbital structure.

The following sections review first the predictions of cosmological
simulations for the structure of the dark halos and baryonic
components in ETGs, and then what has been learnt about their dark
matter halos from different techniques: lensing, hydrostatic hot X-ray
gas athmospheres, and stellar dynamics of halo tracers.

\section{Dark matter halos from cosmological simulations}

\subsection{Dark matter-only simulations}

Simulations of structure formation in a Cold Dark Matter Universe have
progressed greatly over several decades \citep[see][for a detailed
review]{Diemand+Moore11}. For the idealized case of dark matter only,
without baryons, resimulations of individual halos now reach O($10^6$)
particles per halo. The main results from these studies are:

\begin{itemize}

\item Mass density profiles are approximately universal, i.e.,
  independent of halo mass and cosmological parameters aside from
  scalings. The best current density profile model is a 3-parameter
  Einasto profile \citep{NFW96, Bullock+01, Diemand+04, Navarro+10}.

\item Halo concentrations are in the range $c = r_{\rm vir}/r_s \simeq
  5$-$15$ with scatter at given mass $\Delta\log c \simeq
  0.2$. Early-forming halos and smaller halos have higher median
  concentrations \citep{NFW96, Bullock+01, Maccio+08}.

\item Halo shapes are strongly flattened, prolate-triaxial. The mean
  $<c/a> = 0.5\pm 0.1$. The elongation is larger at the center. It is
  approximately stable at fixed radius except in major mergers, and is
  mostly due to anisotropy (median $\lambda\simeq 0.04$)
  \citep{Dubinski+Carlberg91, Jing+Suto02, Allgood+06, Hayashi+07}.

\item Bound substructures are abundant but not dominant, comprising
  $\sim 10\%$ of the halo mass.  They have a steeply declining
  cumulative abundance distribution with mass, $\propto (V_{\rm
    max}/V_{\rm max,host})^{-3}$ or $\propto M^{-2}$ \citep{Klypin+99,
  Gao+04, Springel+08, Diemand+08}.

\item The inner density profiles are cuspy, increasing inwards. The
  asymptotic inner slope is now known to be slightly shallower than
  $-1$ \citep{NFW96, Moore+98, Diemand+04, Navarro+10}.

\end{itemize}

\subsection{Effects of baryons; alignment}

The structure of the simulated DM halos is modified by the
gravitational interaction with the baryonic component, and thus
indirectly by baryonic cooling, settling, star formation and feedback.
The resulting modifications of the halos include:

\begin{itemize}

\item Steepening of the halo density profile: adiabatic contraction
  through slow dissipative settling of gas into a
  disk \citep{Blumenthal+86, Gnedin+04, Sellwood+McGaugh05}.

\item Flattening of the cusp by feedback and shaking \citep{Gnedin+Zhao02,
  Mashchenko+08, Governato+10, Maccio+12}.

\item The halo becomes more spherical and evolves towards oblate at
  all radii but mostly in its inner parts \citep{Kazantzidis+04,
  Bailin+05, Berentzen+Shlosman06, Abadi+10}.

\item Minor axes of the disk and inner halo ($r<0.1r_{\rm vir}$) align
  when a disk forms. The minor axis orientation of the outer halo
  remains unchanged. As a result, the orientations of the inner and
  outer halo are uncorrelated in simulations with baryons, while
  well-aligned in N-body-only simulations \citep{Bailin+05,
    Bailin+Steinmetz05, Deason+11}.

\item The angular momenta of disk and inner halo also align, whereas
  the median misalignment between inner and outer halo angular momenta
  is $\sim25$-$45$ deg \citep{Bett+10, Hahn+10}. Because the halos
  rotate slowly, shape and angular momentum need not be tightly
  correlated. Stacking of simulated halos washes out the alignment
  signal for weak lensing.

\end{itemize}

\subsection{Mass structure of simulated early-type galaxies}

High-resolution models of ETGs in a cosmological
setting are of particular interest for comparing with the empirical
results discussed in later sections of this review.  \citet{Wu+12}
analyzed 42 galaxies with stellar masses of $2.7 \times 10^{10}$-$4.7
\times 10^{11} \msun$ from the cosmological hydrodynamical
resimulations of \citet{Oser+10, Oser+12} which include a model for
cooling, supernova feedback, and star formation processes, but do not
include AGN feedback. These systems contain an in-situ component made
in efficient early star formation, and later grow by accretion through
mostly minor mergers.  

Their stellar density distributions at $z=0$ are approximately
described by Sersic profiles with realistic values of n and $R_e$.  As
measure of their mass distributions, \citet{Wu+12} determined circular
velocity curves (CVCs) $v_c\equiv [GM(r)/r]^{1/2}$.  The dark matter
CVCs are essentially described by power laws outside radii greater
than the softening radius ($0.9 h^{-1}$ kpc for DM particles).  They
vary systematically in that the average slope of the dark matter CVC
increases with increasing circular velocity, or mass, from
approximately flat at low masses to slightly rising ($\simeq 0.3$) at
high masses. When the stellar component is added in, the total CVC
slope changes from slightly falling ($\simeq -0.3$) at low masses to
flat at high masses (see Fig.~1a).  The slopes of dark matter CVCs and
total CVCs correlate with n.  Typical DM fractions in the simulated
ETGs are $\sim20\%$ at $1 R_e$ and $\sim 50\%$ at $5
R_e$. They increase with $R_e$ in kpc,  with n, or with mass
within $R_e$; see \citet{Wu+12}. These results are broadly consistent
with the mass profiles determined from lensing, X-rays and dynamics,
as described in the following.

\section{Dark matter halos of massive elliptical galaxies from lensing
  and X-rays}

\subsection{Halo masses and shapes from weak lensing}

Halo masses for low redshift ($z<0.35$) galaxies are constrained best
by SDSS data \citep[for other work with smaller samples,
see][]{Brainerd+96, Fischer+00, Hoekstra+04, Hoekstra+05, Gavazzi+07}.
The weak galaxy-galaxy lensing signal for 350'000 lenses ($z<0.35$)
from SDSS was compared to a halo model based on simulations ($dn/dM$,
NFW density profile, satellite model) by \citet{Mandelbaum+06a}. 
Satisfactory agreement was found; within this model an $L^*$ galaxy
with $M^*=6\times10^{10} \msun$ is hosted by a halo of mass
$1.4\times10^{12}h^{-1}\msun$.  This value is similar for ETGs in high
and low density regions.  Results from the 2 mag deeper RCS2 survey
already suggested changes with redshift \citep{vanUitert+11}; see
also the talk by Hudson in these proceedings.

In principle, weak lensing is also the method of choice to measure the
outer halo shapes: individual halos will have stronger shear signal on
the projected major axis than on the projected minor axis. However, a
measurement of this effect requires stacking of multiple
galaxies. Thus the signal is preserved only if halos and galaxies are
aligned. Simulations suggest that this is not the case and that the
signal is isotropized; see \S2.  Lensing studies by
\citet{Hoekstra+04, Mandelbaum+06b, Parker+07, vanUitert+12} have
detected differing lensing anisotropy signals. These studies generally
indicate triaxial halos but details are still unclear.

\subsection{Dark matter mass profiles from strong lensing}

Strong lensing \citep[see, e.g.,][]{Treu10} gives an accurate
measurement of the projected mass within the lens Einstein radius,
$R_E$. For lenses at $z\sim 1$, $R_E \sim {\rm several} \, R_e$ and is
located where the DM mass dominates. In this case, the DM can be
studied directly, at least in high-mass ETGs ($\sigma > 200$ km/s)
which dominate the samples.  For most low-$z$ lenses, $R_E < R_e$;
thus the strong lensing analysis measures the combined mass from stars
and DM within $R_E$. Mass profiles can be determined by combining
multiple systems with different $R_E$ \citep{Rusin+Kochanek05,
  Bolton+08, Grillo12}, or by mass reconstruction methods applied to
the source images \citep{Ferreras+05, Leier+11}. The overal result
from these studies is that on average DM starts to dominate at $\sim
2$-$3R_e$.

By combining $M(<R_E)$ with a measured central velocity dispersion or
resolved dispersion profile, it is possible to constrain the average
slope of the mass density profile, $\alpha$. Results from the SLACS
sample show that the density slopes are approximately isothermal
($\alpha=2$) \citep{Treu+Koopmans04, Koopmans+06, Gavazzi+07,
  Barnabe+09, Auger+10}. This is equivalent to a flat CVC in spirals
and has been dubbed the ``bulge-halo conspiracy''.  \citet{Barnabe+09,
  Barnabe+11} also measure halo axis ratios with axisymmetric models,
obtaining values as flattened or rounder than the axis ratios of the
stellar components.  Adopting a Salpeter IMF, the DM fraction inside
$R_E$ is found to increase from 25\% for $\sigma=200$ km/s,
$M^*=10^{11}\msun$ to 70\% for $\sigma=350$ km/s, $M^*=10^{12}\msun$
\citep{Treu10}. Massive ETGs must either be more dark matter
dominated in their cores, or have a heavier, non-universal IMF
\citep{Auger+10, Treu+10}.

\subsection{Hydrostatic mass distributions for X-ray bright ellipticals}

X-ray observations of the hot gas in massive ETGs have long been an
important and independent tracer of their DM halos
\citep{Buote+Humphrey12}. The hot gas extends to large radii where the
DM dominates, and the isotropic pressure makes the analysis
much simpler than for stellar tracers. With {\sl Chandra} and {\sl
  XMM-Newton}, it is no longer a problem to measure temperature
profiles. Thus the current main uncertainties are whether the gas is
in approximate hydrostatic equilibrium, and how large are non-thermal
contributions to the pressure which are not traced by the X-ray data,
such as from relativistic particles, turbulent motions and magnetic
fields.

Some of the main conclusions are as follows:
\begin{itemize}
\item In galaxies with X-ray luminosities $L_X<10^{39}-10^{40}$ erg/s
  (depending on the optical luminosity, often lower mass systems) the
  X-ray flux from the hot gas is subdominant compared to that from
  X-ray binaries and stars, so mass determination is unreliable
  \citep{Revnivtsev+08, Trinchieri+08}.
\item A fraction of the more X-ray luminous systems have strongly
  disturbed X-ray athmospheres \citep{Diehl+Statler07}, so that the
  hydrostatic equilibrium assumption is doubtful.
  \citet{Ciotti+Pellegrini04} quantified how inflows or outflows
  would disturb mass measurements.  Reliable mass determinations are
  only possible in morphologically and dynamically relaxed systems
  \citep{Buote+Humphrey12}.
\item In X-ray bright and relaxed galaxies, the dark matter is
  clearly detected and the inferred potentials are near-isothermal,
  i.e., the CVC is approximately flat to large radii
  \citep{Humphrey+06, Humphrey+12, Churazov+08, Churazov+10,
    Nagino+Matsushita09}. The inferred DM fraction at $2 R_e$ is $\sim
  40$-$80\%$.  \citet{Das+10} using data from \citet{Churazov+10} found
  that the CVCs turn up at large radii, possibly due to the
  surrounding group DM halo, and that the typical circular
  velocity $v_c(R_e)$ correlates with the velocity dispersion of the
  sub-Mpc group environment.
\item Comparing determinations of the gravitational potential or CVC
  for the same galaxy from X-ray and optical observations gives a way
  to estimate the non-thermal contributions to the total
  pressure. Values found are between $\sim 10\%$ in quiescent galaxies
  like NGC 1399 to $\sim 30\%$ in more disturbed systems like M87
  \citep{Churazov+08, Churazov+10, Das+11}, with large uncertainties
  because of possible systematic effects in both techniques.
\end{itemize}

\section{Dark matter halos from dynamics}

\subsection{Kinematic halo tracers}

Traditional long-slit kinematics reaches down to surface brightnesses
of $\mu_V \sim 23.5$ corresponding to $R\sim2 R_e$
\citep[e.g.][]{Kronawitter+00}.  To determine the DM mass
profile and the halo orbit distribution, alternative data are therefore
needed which reach to larger radii and fainter surface brightnesses,
such as
\begin{itemize}
\item Planetary nebulae (PNe) velocities measured from their bright
  [OIII] emission line \citep[e.g.][]{Hui+95, Arnaboldi+96, Mendez+01,
    Peng+04}.  PNe are generally good tracers of the stellar light and
  kinematics in ETGs \citep{Coccato+09}.  Useful kinematic information
  can typically be derived to $\sim 5$-$8 R_e$, and up to beyond 100 kpc
  in some cases \citep{Doherty+09}, reaching $\mu_V \sim 27.5$.
\item Globular cluster (GC) velocities measured from their absorption
  line spectra \citep[e.g.][]{Cote+01, Hwang+08, Schuberth+10, Woodley+10}.
  Especially bright, central ETGs contain large GC systems reaching to
  very large radii.  The use of GCs as mass tracers is somewhat
  complicated by the fact that they do not trace the galaxy light
  directly, requiring simultaneous determination of their
  (completeness-corrected) number density profile.
\item Stacked absorption line spectra from integral field units (IFU)
  placed at large galactocentric radii. This method has been employed
  using the Sauron \citep{Weijmans+09} and VIRUS-P \citep{Murphy+11}
  IFU instruments and reaches down to $\mu_V\sim25.5$.
\item Stacked absorption line spectra from off-target pixels in
  slitlets originally placed in the halo to measure globular cluster
  velocities \citep{Proctor+09, Foster+11}, reaching $\mu_V\sim25$.
\end{itemize}

\subsection{Outer halo kinematics from planetary nebulae}

While in early work with 4m telescopes velocities could be measured
only for the brightest $\sim 30$-$50$ PNe in ETGs at Virgo distance
\citep[e.g.][]{Arnaboldi+96, Arnaboldi+98}, we now have $\sim 20$ ETGs
with hundreds of measured PN velocities for tracing the velocity field
out to large radii, typically $5$-$8 R_e$.  This is mostly thanks to
slitless spectroscopy efforts at the VLT \citep{Mendez+01,
  Teodorescu+05, Teodorescu+11, McNeil+10, McNeil-Moylan+12} and to
data from the special-purpose PN.S instrument \citep[see][]{Douglas+07,
  Coccato+09, Napolitano+09, Napolitano+11}. The kinematic properties
of ETG halos derived from these PN samples include
\citep[see][]{Coccato+09}:
\begin{itemize}
\item The ratio of mean velocity ${\bar v}$ and dispersion $\sigma$ in
  the halo correlates with that within Re for the greater part of the
  sample, but some galaxies are more rotationally dominated in their
  halos.
\item The division into slow and fast rotators defined by the
  kinematics within $\sim R_e$ \citep{Emsellem+07}, is largely
  preserved in the outer halos. A few galaxies have more complex
  profiles of the angular momentum-related $\lambda_R$ parameter.
\item Kinematic misalignments appear to be more frequent in the halos.
\item The radial profiles of rms velocity ${\bar v}^2+\sigma^2$ fall
  into two groups, the major group characterized by a slow radial
  decrease, and a second group with ``quasi-Keplerian'', steeply
  falling profiles, first drawn attention to by \citet{Mendez+01} and
  \citet{Romanowsky+03}.
\end{itemize}

\subsection{Kinematic substructure}

Cosmological galaxy assembly models predict that massive galaxies grow
by accreting smaller companions. During and after minor merger events
photometric and kinematic substructure would be visible in the outer
halo of the host galaxy where dynamical time-scales are long. This is
an interesting subject (and opportunity) by itself, but is also
relevant for dynamical mass determinations at large radii.  One
example is the nearby BCG galaxy NGC 1399, studied by
\citet{McNeil+10} with PNe and by \citet{Schuberth+10} with GCs.  The
phase-space plot for PNe around NGC 1399 in Fig.~1b shows three
components, including PNe from NGC 1399 around $V_{1399}\simeq 1400
{\rm km/s}$, PNe from the lower-luminosity ETG NGC 1404 close-by in
projection, and finally PNe from a component of ``low-velocity
outliers'' with velocities $\sim 750\pm250 {\rm km/s}$.

The non-equilibrium low-velocity component is also seen in both the
blue and red globular clusters around NGC 1399.  The number density
profile of red GCs approximately follows stellar light in this galaxy,
whereas that of blue GCs does not. The decontaminated system of red
GCs appears to be in approximate dynamical equilibrium, but this seems
not the case for the blue GCs \citep{Schuberth+10}.  This could be
related to the fact that blue GCs are particularly prominent in
smaller galaxies \citep{Strader+06} which would be preferentially
accreted at late times.

Indications for substructure have been found also in the halos of M87
and NGC 4649 \citep{Romanowsky+12, Coccato+12}.  A fraction of large
ellipticals have a second galaxy within a 10-20’ field.  With discrete
tracers (PNe or GCs) it is possible to eliminate kinematic
substructures, which is a prerequisite for a reliable dynamical mass
determination in such systems.

\subsection{Dark halo circular velocity curves and densities from stellar kinematics}

Using absorption line kinematics to 1-$2 R_e$ for luminous round ETGs,
and nonparametric spherical distribution function models in combined
luminous and dark halo potentials, \citet{Gerhard+01} found that the
CVCs of ETGs are approximately flat.  A more recent analysis of a
sample of ETGs in the Coma cluster with M$_B=$[-18.8,-22.6] and
kinematic data out to 1-$3R_e$ with axisymmetric Schwarzschild models
by \citet{Thomas+07} showed somewhat more varied CVC shapes. Massive
ETGs tend to have flat CVCs \citep[e.g., NGC 4649,][]{Das+11}, while
lower-mass ETGs (see \S4.5) are also consistent with slightly
decreasing CVCs. These studies agree in their derived DM
fractions, $\sim 10-40\%$ within $R_e$, assuming constant M/L for the
stellar component, which is also consistent with models of inner 2D
kinematics by \citet{Cappellari+06}.

Both studies also agreed in finding significantly higher central dark
matter densities in ETGs than in spiral galaxies of the same
luminosity or mass. For the Coma ETGs, \citet{Thomas+09} found a factor of
$7\times$ higher mean DM density within $2R_e$ at the same stellar
mass, and $13\times$ at the same luminosity. Baryonic contraction is
not sufficient to explain this difference.  One simple
explanation is that the cores of ETG halos assembled earlier (at
redshifts 1-3) than spirals of same luminosity \citep{Gerhard+01,
  Thomas+09}. Good agreement of the dynamical luminous plus dark
matter models was found with SLACS lensing models of \citet{Auger+09};
see \citet{Thomas+11}.

\subsection{Mass Distributions of Quasi-Keplerian Ellipticals}

\citet{Mendez+01} first showed that the steeply decreasing outer
velocity dispersion profile of the intermediate-luminosity elliptical
(ILE) galaxy NGC 4697 could be matched well with an isotropic
Hernquist model. They thus concluded that no evidence for dark matter
had been found out to $3R_e$, but that dark matter could be present if
the velocity distribution is anisotropic. \citet{Romanowsky+03} drew
attention to three further ILEs with steeply decreasing $\sigma_p(R)$,
NGC 821, 3379, and 4494, suggesting the presence of little if any dark
matter in their halos. However, the well-known mass-anisotropy
degeneracy \citep{Binney+Mamon82} is much stronger in galaxies such as
these ILEs, with de Vaucouleurs type luminosity profiles and steeply
decreasing $\sigma_p$-profiles, than in galaxies with either more
shallow luminosity profiles or radially constant projected velocity
dispersions \citep{Gerhard93}.

With NMAGIC particle models \citep[see][]{deLorenzi+07} based on a
variety of kinematic data, including several hundred PN line-of-sight
velocities, \citet[][for NGC 4697]{deLorenzi+08}, \citet[][for NGC
3379]{deLorenzi+09} and \citet[][for NGC 4494]{Morganti+12} determined
the range of quasi-isothermal halo potentials consistent with the data
for these galaxies. Fig.~1c shows the 70\% confidence boundaries from
the PN likelihood for NGC 3379 and 4697, and from a more comprehensive
analysis for NGC 4494. While some spread in the allowed CVCs and
anisotropies remains, as expected, dark matter is required in all 3
galaxies. Fig.~1c indicates that the most favoured CVCs may be
slightly falling, as in the lower-mass systems in \citet{Wu+12}.
Relative to the outer CVCs, the baryonic centers of these galaxies are
quite centrally concentrated, as could be expected in gas-rich
mergers.

\begin{figure}[hp]
\begin{center}
\resizebox{6.5cm}{!}{\includegraphics{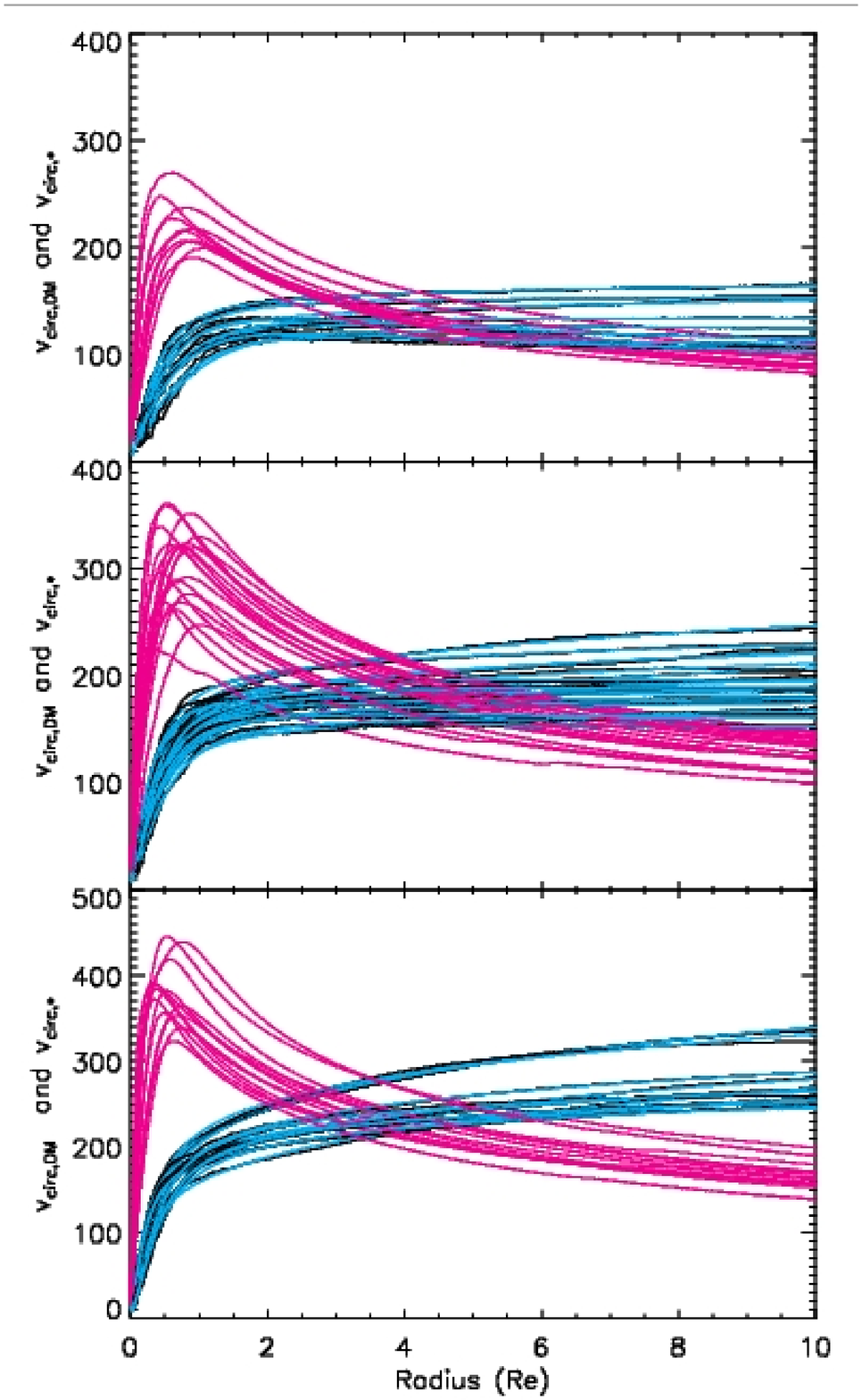}}\resizebox{6.5cm}{!}
                   {\includegraphics{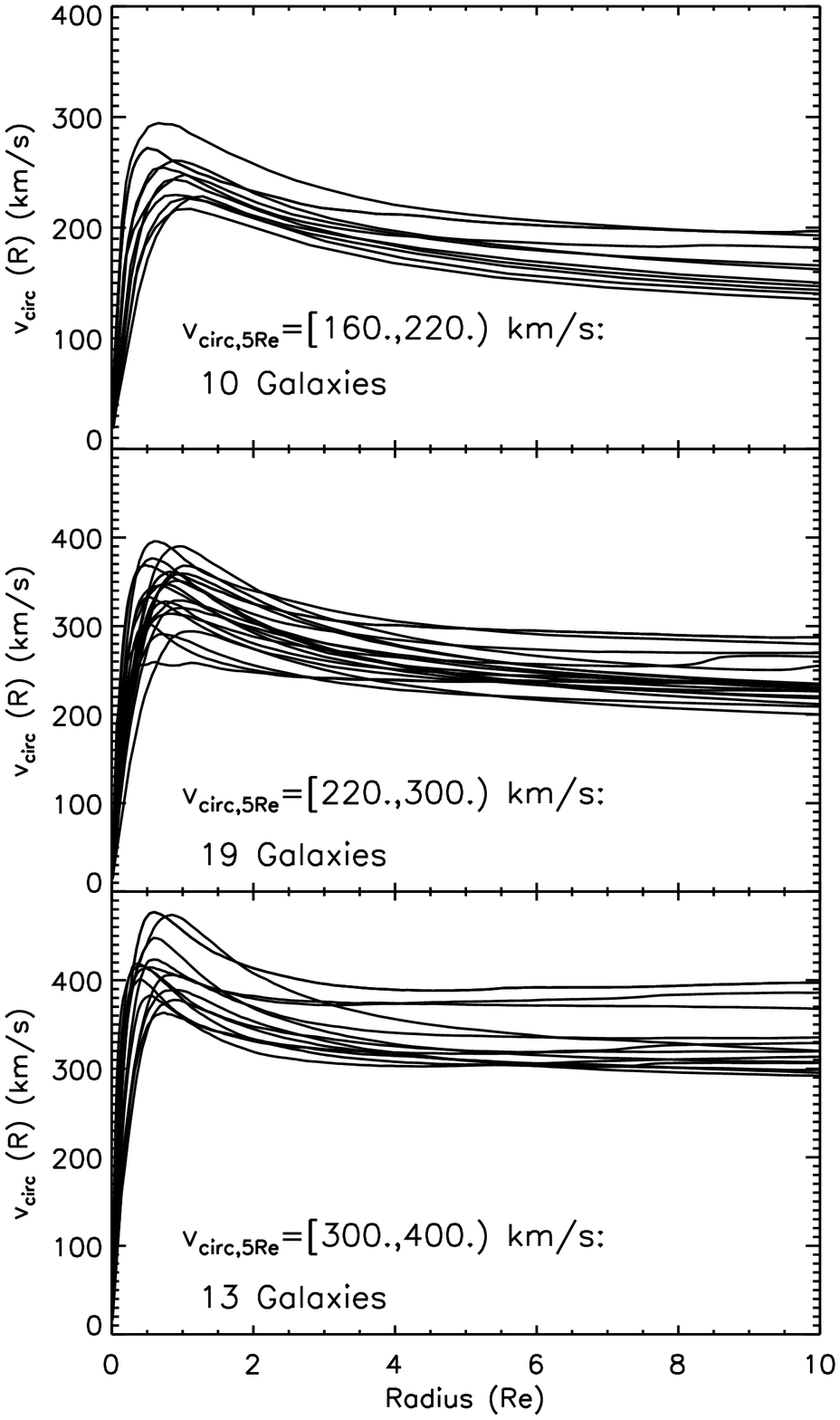}}
\resizebox{5.5cm}{!}{\includegraphics{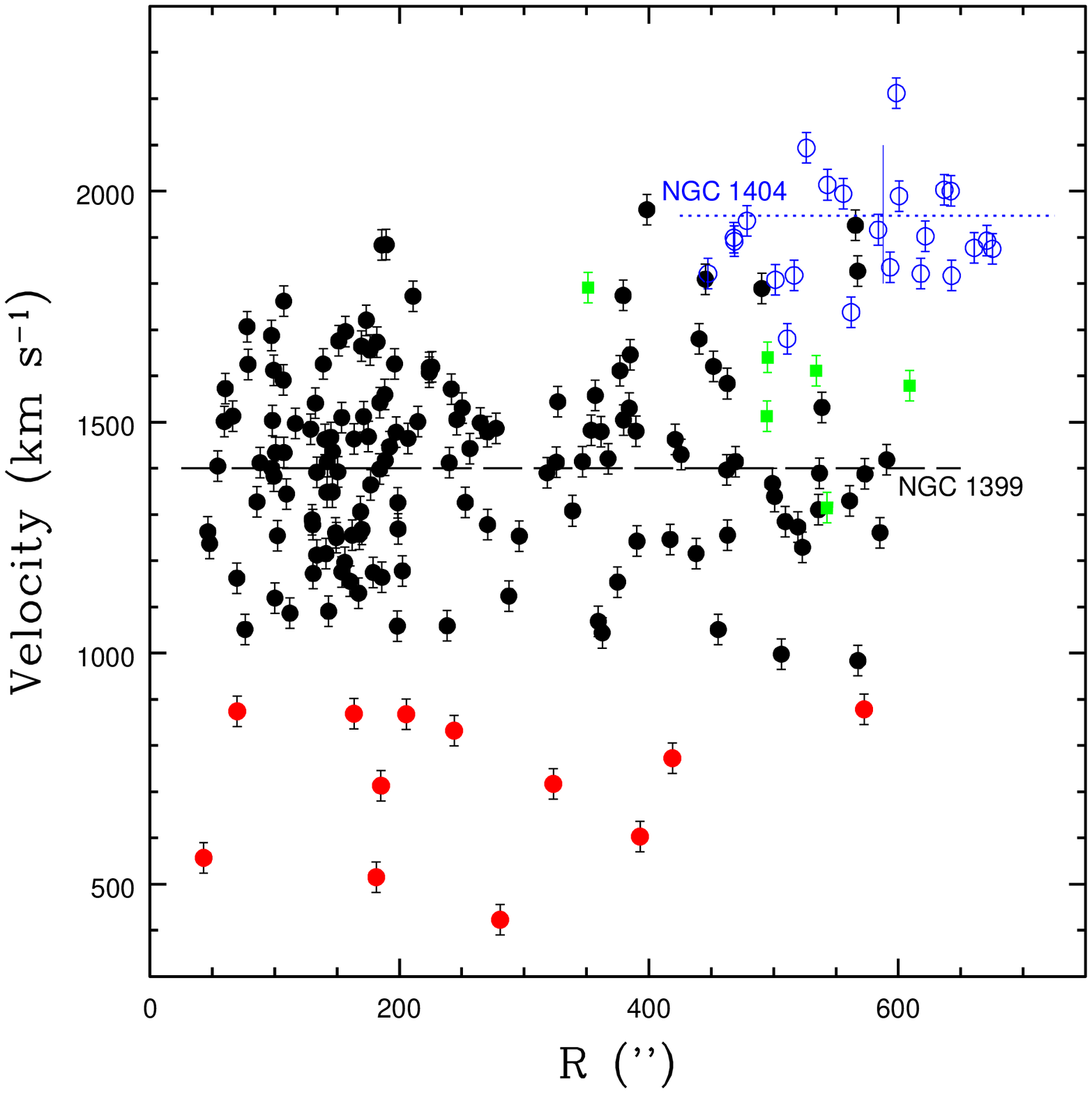}}\resizebox{7.5cm}{!}
                   {\includegraphics{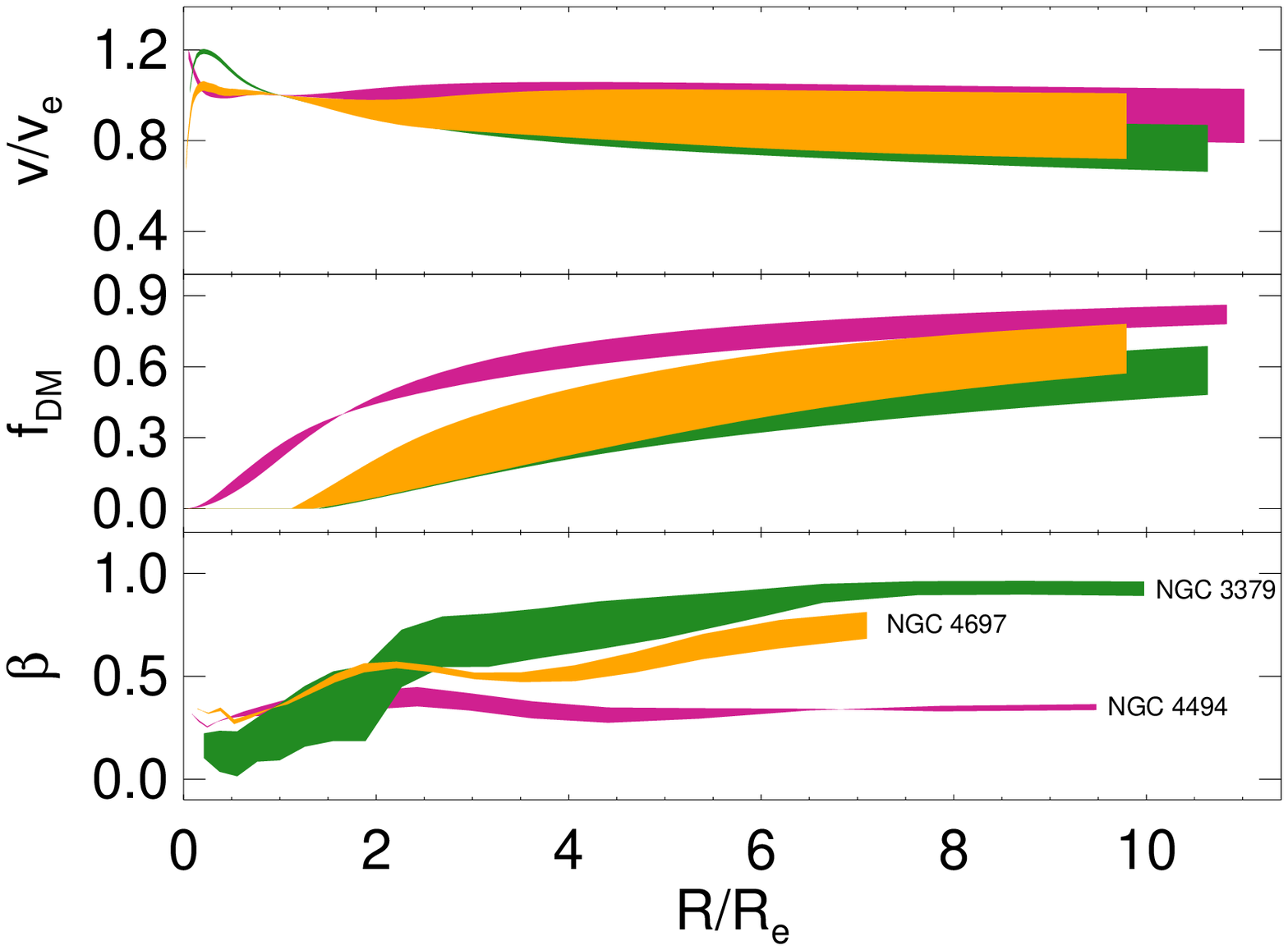}}
\vspace*{0.5 cm}
\caption{(a) Top: Circular velocity curves from stars (magenta), dark
  matter (black-blue, both left panel) and total mass distribution
  (right), for 42 model galaxies from the cosmological simulation of
  \citet{Oser+12}, in bins of increasing total circular velocity at
  $5R_e$, as given on the right panels.  From \citet{Wu+12}.  (b) Bottom
  left: Phase-space distribution of PNe in the nearby BCG galaxy NGC
  1399 (black), its neighbour galaxy NGC 1404 (blue), and the
  low-velocity component (red). From \citet{McNeil+10}.  (c) Bottom right:
  70\% confidence ranges from dynamical modeling for total circular
  velocity normalized to value at $R_e$, DM fraction, and
  anisotropy parameter, for the three quasi-Keplerian ETGs NGC 3379,
  4494, 4697. From \citet{Morganti+12}. }
   \label{fig1}
\end{center}
\end{figure}

\subsection{Dark Matter Fraction with Mass}

For radially constant mass-to-light ratio $\Upsilon$, maximizing the
fraction of mass following the luminosity profile results in a lower
limit for the DM fraction, $f_{\rm DM, min}$. Dynamical modeling
(\S4.4, 4.5) gives $f_{\rm DM, min}(R<1R_e)\simeq0.1$-$0.4$ and
$f_{\rm DM, min}(R<5R_e) \simeq 0.3$-$0.8$, with a tendency of larger
values for more massive ETGs.  The corresponding $\Upsilon_{\rm max}$
also increase with galaxy mass \citep[see][]{Thomas+11}.
\citet{Napolitano+05} estimated gradients of $\Upsilon$ for a sample
of ETGs with available dynamical mass modeling and found a correlation
between luminosity and $\Upsilon$-gradient such that the more luminous
galaxies were more dark matter dominated. Comparing with DM halo
predictions, the underlying reason appeared to be that the stellar
bodies of brighter galaxies enclose a larger fraction of their DM
halos.  \citet{Deason+12} used power-law models to estimate dynamical
masses at $5R_e$ for a sample of ETGs with extended PN and GC tracer
kinematics. From their models they also found an increase of DM
fraction inside $5R_e$ with galaxy mass, independent of whether a
Chabrier or Salpeter IMF was assumed. The strong lensing results
(\S3.2) show a similar trend for the DM fraction within $R_E$.

Because dynamical $\Upsilon_{\rm max}$ increase with galaxy mass
faster than stellar population $\Upsilon$ for given IMF, either some
of the dark matter in massive ETGs must follow the light distribution,
or the IMF must vary systematically with galaxy mass
\citep{Thomas+11}. The latter interpretation is favoured by the
dynamical analysis of \citet{Cappellari+12}, and by new stellar
population modeling of ETG spectra with NIR spectral features
sensitive to low mass stars \citep{Conroy+vanDokkum12}.  If the IMF
varies between galaxies, it is natural in a hierarchical assembly
picture to expect that it would also vary with radius within a
galaxy. Then reliable mass decomposition in ETGs will require radially
resolved spectral measurements of the IMF.

\section{Conclusions}

\begin{itemize}

\item Pure N-body simulations in $\Lambda$CDM cosmology predict nearly
  universal, prolate-triaxial DM halos.  Including baryons and galaxy
  formation modifies the DM halo shapes, alignment, and inner density
  profiles.  This provides one way to constrain the formation
  processes of the baryonic components.

\item Weak lensing measurements are consistent with the predicted
  density structure of halos on large scales, and determine the
  relation between halo mass and galaxy luminosity.

\item Mass determinations from strong lensing, X-ray emitting gas,
  stellar kinematics and PN kinematics out to several $R_e$ find flat,
  close to isothermal CVCs for massive ETGs, while for lower mass,
  ``quasi-Keplerian'' ETGs the CVCs may be slightly falling.

\item DM fractions at $5 R_e$ inferred from dynamical modeling are in
  the range $\sim 30$-$80\%$.  Inner DM densities in ellipticals are
  $\sim 10$ times higher than in spiral galaxies of the same stellar
  mass.

\item A suite of recent high-resolution cosmological simulations of
  massive galaxies grown by minor and major mergers show systematic
  variations of the CVCs and DM fractions inside $5R_e$ with
  mass and fitted Sersic index, similar to those observed in ETGs.

\end{itemize}




\end{document}